# Time-resolved four-wave-mixing spectroscopy for inner-valence transitions

Thomas Ding,[1, *] Christian Ott,[1,2] Andreas Kaldun,[1] Alexander Blättermann,[1] Kristina Meyer,[1] Veit Stooß,[1] Marc Rebholz,[1] Paul Birk,[1] Maximilian Hartmann,[1] Andrew Brown,[4] Hugo Van Der Hart,[4] and Thomas Pfeifer,[1,3]

[1] Max-Planck-Institut für Kernphysik, Saupfercheckweg 1, 69117 Heidelberg, Germany
[2] Chemistry Department, University of California, Berkeley, California, 94720, USA
[3] Centre for Quantum Dynamics, Ruprecht-Karls-Universität Heidelberg, 69120 Heidelberg, Germany
[4] Centre for Theoretical Atomic, Molecular and Optical Physics, Queen's University Belfast, Belfast BT7 1NN, UK

**Non-collinear four-wave mixing (FWM) techniques at near-infrared (NIR), visible, and ultraviolet frequencies have been widely used to map vibrational and electronic couplings, typically in complex molecules. However, correlations between spatially localized inner-valence transitions among different sites of a molecule in the extreme ultraviolet (XUV) spectral range have not been observed yet. As an experimental step towards this goal we perform time-resolved FWM spectroscopy with femtosecond NIR and attosecond XUV pulses. The first two pulses (XUV-NIR) coincide in time and act as coherent excitation fields, while the third pulse (NIR) acts as a probe. As a first application we show how coupling dynamics between odd- and even-parity inner-valence excited states of neon can be revealed using a two-dimensional spectral representation. Experimentally obtained results are found to be in good agreement with *ab initio* time-dependent R-matrix calculations providing the full description of multi-electron interactions, as well as few-level model simulations. Future applications of this method also include site-specific probing of electronic processes in molecules.**

With the development of coherent femtosecond-duration laser pulses nonlinear (third-order) four-wave-mixing (FWM) spectroscopy has become a versatile tool for the investigation of ultrafast dynamics in molecules and other material samples. A large body of experimental approaches for time-resolved molecular spectroscopy has been developed based on FWM phenomena, such as photon echo [1], coherent anti-Stokes Raman scattering [2], transient grating [3], and many more. The spectroscopically most comprehensive implementation of FWM is the so-called two-dimensional spectroscopy (2DS) with three temporally independent pulses which allows the correlation of excitation– and nonlinear-response spectra to directly measure couplings between quantum states [4,5]. 2DS in the infrared, visible, and ultraviolet spectral ranges has enabled the exploration of vibrational [6], electronic [7] and vibronic [8] coupling dynamics in complex molecular systems. The extension of 2DS into the extreme ultraviolet (XUV) and X-ray spectral regions aims to map coupling dynamics between spatially localized inner-valence or core transitions among different sites of a quantum system. It has been theoretically discussed by Mukamel and coworkers [9,10] and is a long-awaited goal. In principle, modern free-electron laser (FEL) and high-harmonic generation (HHG) based coherent XUV and X-ray light sources can provide appropriate pulses for 2DS experiments, but key challenges are both the increased technical demands to create appropriate multi-pulse geometries in this energy range, as well as the intrinsically low photon flux of laboratory based sources. Nevertheless, first steps are being made, as demonstrated in recent experiments that explore NIR- and XUV-induced transient gratings that emphasize, respectively, both the spectroscopic element selectivity [11] and the enhanced spatial resolution [12] capabilities of the XUV domain.

Meanwhile, an all-optical two-color pump-probe technique utilizing weak attosecond XUV pulses and strong few-cycle NIR pulses—often referred to as attosecond transient absorption spectroscopy (ATAS)—has opened a direct route to the measurement and control of the XUV spectral response of bound electronic transitions [13–21]. With both the XUV-NIR time delay $\tau$ and the NIR intensity $I_{NIR}$ as continuously tunable control parameters, ATAS has a multi-dimensional experimental anatomy [22]. Two-dimensional (2D) time-domain spectra $S(\tau, \omega)$ measured as a function of both the XUV photo-excitation frequency $\omega$ and time delay $\tau$ exhibit a characteristic fringe pattern across coherence resonances regardless of the specific target system [23]. The Fourier transform along the signal's pulse-delay time trace is taken ($\tau \rightarrow \nu$) to decompose the interference pattern into a two-dimensional absorption spectrum (2DAS), $\tilde{S}(\nu, \omega)$, which exhibits diagonal regions with point-like and/or line-like peaks [13,18,22,24,25] revealing information about the dynamical pathways along which the system is driven. We recently introduced an analytical framework to understand and characterize the signatures of time-delay dependent polarization dynamics in 2D-XUV absorption spectra [23].

Thus far, as in any one-photon-absorption-based technique, the spectroscopic information provided by ATAS has been restricted to the dipole-allowed XUV-transitions. Accordingly, the interference structures observed in the 2DAS only consisted of maxima at zero or even multiples of the NIR-photon energy. Interferences at odd multiples of the NIR-photon energy would indicate couplings to additional (one-photon forbidden) channels, which could be accessed by a combination of XUV and NIR excitation fields, as previously applied to photo-electron spectroscopy [26,27].

Here we extend ATAS into a time-resolved FWM technique for the spectroscopy on both dipole-allowed and –forbidden XUV-excited inner-valence transitions by adding a second weak NIR pulse to the above described ATAS geometry. The first two pulses (XUV and weak NIR) coincide in time (with locked zero-delay timing and a fixed phase relation between fundamental field and higher harmonics) and are temporally separated from the strong (non-perturbative) third (NIR) pulse by the time delay $\tau$. The interaction of a sample with the first two pulses (XUV and NIR) creates a polarization, i.e., the coherent superposition of ground and both odd- and even-parity excited states. This two-color pump step extends the coherent excitation onto states the transition into which would be forbidden by a single XUV photon from the ground state. After the excited system has evolved freely during the time delay $\tau$, the interaction with the strong NIR pulse generates a nonlinear (third-order) response signal by again coupling between XUV-dipole-allowed and –forbidden excited states. We have chosen the $2s$-inner-valence excited states $2s^{-1}3s(^1S^e)$, $2s^{-1}3d(^1D^e)$ and $2s^{-1}3p(^1P^o)$ of neon (henceforth denoted as $3s$, $3d$, and $3p$) as a target to perform first such proof-of-principle FWM experiments. Neon provides an appropriate electronic energy-level structure for investigating resonant transitions between XUV-dipole-allowed and –forbidden states which cannot be accessed by conventional ATAS methods.

The experimental setup (see Fig. 1a), as previously described in Ref. [17], involves a commercial Ti:Sapphire multipass amplifier with hollow-core fiber and chirped-mirror compression stages for the generation of NIR (central photon energy ~1.6 eV) sub-7 fs pulses delivered at a 4 kHz repetition rate and ~0.4 mJ pulse energy. These pulses act as driver pulses for HHG in an argon gas-filled cell (~75 mbar backing pressure) yielding short trains of attosecond pulses that are inherently phase-locked to the fundamental field. HHG spectra obtained extend continuously over the autoionizing resonance region of neon between 43 eV and 50 eV covering the full $2s^{-1}np$ Rydberg series (cf. the HHG absorption spectrum in Fig. 1b). A piezo-driven two-component split mirror (inner, dynamic part: gold coating; outer part: silver coating) is employed to introduce a time delay between the co-propagating attosecond XUV and femtosecond NIR pulses. The NIR pulse intensity is controlled by a piezo-controlled iris aperture. Spatial beam separation between XUV and NIR is achieved by a two-part spectral band-pass filter in annular geometry where the outer, annular, part consists of a 2 µm thin nitrocellulose membrane (1.2 mm central hole) and the inner part is a single layer of 0.2 µm thin aluminum foil (1.2 mm diameter). Importantly, for the results shown in this work, the central part transmits a residual NIR-photon intensity on the order of $1-10\%$, which can be considered as a perturbative NIR replica pulse which remains locked in temporal overlap with the XUV. The NIR transmission is due to small holes or other imperfections in the extremely fragile foil (e.g. created during the manufacturing process and/or mechanical strain while assembling the custom-made filter arrangement). After passing the band-pass filter, the NIR-pulse delay (with respect to the NIR-XUV excitation step) and intensity-controlled laser beams are focused into the neon gas-filled target cell (~35 mbar backing pressure). The transmitted XUV light contains the sample's dipole response which is detected by a high-resolution XUV spectrometer (~20 meV Gaussian standard deviation near 45 eV, see Ref. [16] for details).

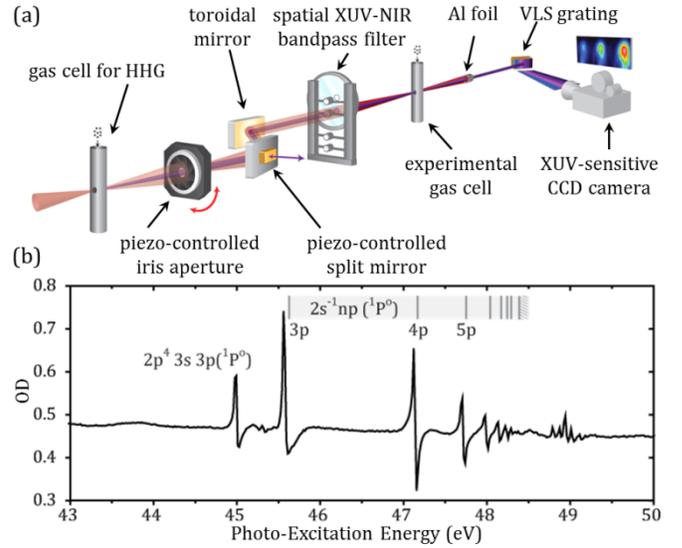

**Fig. 1.** (a) Schematic of the experimental setup and the key elements, as discussed in the main text. Essential for creating the described pulse configuration for FWM (temporally locked XUV-NIR excitation fields and NIR probe field) is the piezo-controlled split mirror in combination with an annular XUV-NIR filter geometry transmitting a residual NIR-photon flux through the inner Al filter. (b) Linear spectroscopy (XUV only): The natural neon line spectrum in terms of optical densities (OD). Resonances are assigned based on Ref. [38].

For comparison with our experimental results, calculations based on the R-matrix with time-dependence (RMT) approach [28] were performed for the neon atom. RMT is an *ab initio*, non-perturbative computational technique for the description of general multi-electron atomic/ionic systems interacting with a strong laser field. By employing the standard R-matrix division of space [29] a tractable solution of the time-dependent Schrödinger equation can be obtained affording a full account of multi-electron effects, and a comprehensive treatment of detailed atomic structure. In order to describe both the core– and doubly-excited states of neon under investigation, the calculations comprise all single–, double– and triple-excitations of the $2s$ and $2p$ electrons into $3p, 3s, 3d$, (spectroscopic) $\overline{4s}$ and $\overline{4p}$ (pseudo)-orbitals, which are determined as described in Ref. [30]. This leads to an expansion over 152 multi-electron configurations. Additionally, to account for the various ionization pathways, six residual ion states of $Ne^+$ are included in the calculation. This yields energies within 0.08 eV of the literature values, for the states under investigation [31–33]. Using RMT we compute $d(t)$: the time-dependent expectation value of the dipole operator. The absorption spectrum is then given by [34]

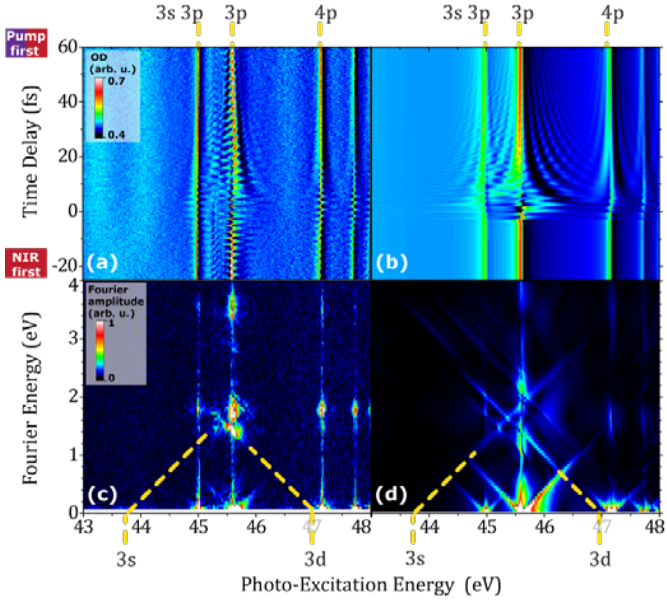
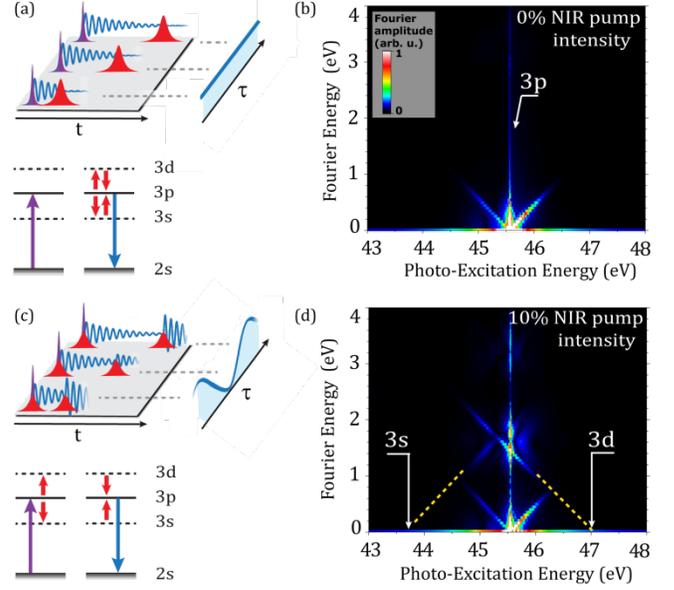

**Fig. 2.** Time-resolved 2D absorption spectra $S(\tau,\omega)$ obtained from (a) experiment and (b) full *ab initio* calculations at ~$10^{12}$ W/cm² NIR-field intensity (also see Dataset 1 in Ref. [43]). Strong periodic full-cycle modulations are imprinted on the $3p$ resonance which gives rise to line-like peaks at ~1.6 eV Fourier energy in the corresponding 2DAS, $\tilde{S}(\nu,\omega)$ (lower panels c and d, respectively), indicating coherent coupling pathways. The energetic position of the dipole-forbidden and thus spectroscopically hidden $3s$ and $3d$ coupling partners ($\omega_{3s}$ = 43.7 eV and $\omega_{3d}$ = 47.0 eV) are indicated by the dashed lines as a guide to the eye. The $3s$ and $3d$ resonances are assigned based on Ref. [32].

$$\sigma(\omega) = 4\pi\frac{\omega}{c}\,\text{Im}\left[\frac{d(\omega)}{E(\omega)}\right] \quad (1)$$

where $\omega$ is the photon energy, $c$ is the speed of light, and $d(\omega)$ and $E(\omega)$ are the Fourier transforms of $d(t)$ and the electric field, respectively.

In Figs. 2(a) and 2(b), we present time-delay-resolved absorption measurements for a scanned time delay $\tau$ between XUV-NIR pump and NIR-probe fields, obtained from experiment and RMT calculations. Time-resolved absorption changes can be observed across all the resonance lines shown in Figs. 2(a) and 2(b). Our focus here is on the absorption changes occurring near the $3p$ state at $\omega_{3p}$ = 45.55 eV photo-excitation energy in order to understand the two-photon XUV-NIR-induced wave-packet dynamics created by the superposition of two coherently excited states with different parities. The time-domain absorption spectra $S(\tau,\omega)$ (Figs. 2a and 2b) are Fourier analyzed ($\tau \to \nu$) to decompose the temporal beat patterns for each resonance into distinct peaks on a 2DAS, $\tilde{S}(\nu,\omega)$, spectral representation (Figs. 2c and 2d). Peak position, shape, and orientation allow the characterization of coherent coupling and transition pathways analogous to the case previously described for XUV-only excitation and a single time-delayed NIR probe pulse [23]. For the case presented here, the additional weak NIR pulse locked to the XUV excitation pulse near time zero plays a crucial role, as it coherently excites states that would otherwise be dipole-forbidden for an XUV-only excitation. The basic structures observed in the time-domain spectra $S(\tau,\omega)$ (Figs. 2a and 2b) have been seen in several related ATAS studies with XUV-only-excitation fields [13–19,22,23,34–37] and can in

**Fig. 3**. Schematic illustrations of the $\tau$-dependent third-order coupling mechanisms (panels a and c) and corresponding few-level model simulations (panels b and d). (a) The XUV pulse excites an $2s$ electron (violet arrow) creating a coherent superposition of the ground state and the $3p$ state while XUV transitions into final states $3s$ and $3d$ are dipole-forbidden. The superposition leads to coherent dipole emission (blue wavy lines). After the time delay $\tau$ the NIR pulse induces transient coupling (red arrows) which transfers population out of the $3p$ into $3s$ and $3d$ states. Amplitude and phase modifications lead to the characteristic spectral features near zero Fourier Energy in the 2DAS (panel b). (c) Two-photon (XUV-NIR) excitation into a coherent superposition of $3p$, $3s$, and $3d$ states. After the time delay $\tau$, during which the superposition has freely evolved in time, the interaction with the third (NIR) pulse mediates the exchange of population between these states which leads to constructive or destructive dipole emission of the $3p$ state periodically with the optical period of the NIR pulse as a function of time delay. In the 2DAS (panel d) this creates the line-like features near a Fourier energy corresponding to the NIR photon energy.

general be decoupled into two interfering sub-structures [23]: $(i)$, a slow, $\tau$-dependent modulation following a hyperbolic geometry near resonance lines corresponds to a fork-like feature with elliptically-shaped peaks with slope 1 originating at zero Fourier energy in the 2DAS $\tilde{S}(\nu,\omega)$. $(ii)$, a fast $\tau$-dependent rippling-type pattern, which gives rise to line-like Fourier peaks of slope 1 at non-zero Fourier energies. According to dipole-selection rules such Fourier peaks at even-numbered multiples of the NIR-photon energy indicate transitions between identical-parity states mediated by an even number of NIR photons. Such features (half- and quarter-cycle modulations) have been observed and interpreted recently [13,18,22]. In Figs. 2(c) and 2(d) we observe these Fourier features, however, at one NIR-photon energy, ~1.6 eV (full-cycle modulations), evidencing a coherent excitation and time-dependent coupling of opposite-parity states. We identify the even-parity $3s$ and $3d$ states as the coupling partners of the $3p$ state by analyzing the orientation of the two dominant line-like peaks appearing at $\omega_{3p}$ = 45.55 eV photo-excitation energy and ~1.6 eV Fourier energy, which extrapolate towards the coupled resonance energies on the photo-excitation energy axis at $\omega_{3s}$ = 43.7 eV and $\omega_{3d}$ = 47.0 eV [32]. It is important to note that the two-color XUV-NIR pump step for the coherent excitation of both

XUV-dipole-allowed and dipole-forbidden (allowed by the absorption of an additional NIR photon near time zero) transitions to final states $3p$ and $3s,d$ is the key mechanism to observing Fourier features at odd-numbered multiples of the NIR-photon energy. To further confirm this understanding we present in Fig. 3 few-level model (FLM) simulations based on the approach previously described in Ref. [16]. The numerical parameters (resonance energies, line widths, asymmetry $q$-parameters) used to model the three states ($3p, 3s, 3d$) above the ground state are taken from Refs. [32,38]. The $N = 1$ ionization continuum is not included but we account for line-shape asymmetry effects due to continuum coupling [39] by the phase shift $\varphi(q) = 2\arg(q - i)$ of the time-dependent dipole response function [17]. The dipole-matrix elements were estimated based on hydrogen-like orbitals including a semi-empirical shielding of the nuclear charge by the partially filled inner shell [40]. We note the good agreement of the FLM simulation (Fig. 3d) with both the experimental and the RMT spectra shown in Fig. 2 which further substantiates our understanding of the underlying mechanism as a third-order (four-wave mixing) interaction process with three excited resonances ($3p, 3s, 3d$). In Fig. 3 we present a schematic view of the discussed mechanism and FLM simulation results assuming a NIR-pump-intensity of 10 % of the fundamental, non-perturbative NIR pulse, respectively. In principle, this method allows the extraction of coupling strengths (dipole-matrix elements) from the experimental data, as soon as the pulse shape and intensity of the NIR pulse can be accurately determined [41].

In conclusion, we have demonstrated the measurement of the ultrafast nonlinear response from 2$s$-inner-valence excited states in neon by applying three short-pulsed laser fields (XUV-NIR-NIR) in an ATAS-type FWM geometry. This enables a direct access to coupling dynamics between states of opposite parity which is not possible with conventional two-pulse ATAS techniques. The FWM mechanism that underlies these dynamics was interpreted and fully understood on the basis of few-level model simulations. To account for the electronic correlation of the neon atom in full dimensionality we compared our experimental results with RMT calculations and found excellent agreement. This renders the presented methodology a promising experimental tool to investigate polarizabilities in multi-electron systems [42] even in excited and metastable states of different parity on ultrafast timescales. Additionally, the method can be extended to more complex systems, including the site-specific probing of intramolecular electron dynamics and excitation transfer.

**Funding:** Deutsche Forschungsgemeinschaft (DFG) (PF 790/1-1); European Research Council (ERC) (X-MuSiC-616783); U.K. EPSRC (EP/G055416/1) CORINF under the Marie Curie Action of the European Commission. This work used the ARCHER UK National Supercomputing Service.